# FAIR evaluation of ten widely used chemical datasets: Lessons learned and recommendations

**Marcos Da Silveira   Oona Freudenthal    Louis Deladiennee**
Luxembourg Institute of Science and Technology
5, avenue des Hauts-Fourneaux, L-4362 Esch-sur-Alzette, Luxembourg
*Contact : marcos.dasilveira@list.lu*

## Introduction

This document focuses on databases disseminating data on (hazardous) substances found on the North American and the European (EU) market. The goal is to analyse the FAIRness (Findability, Accessibility, Interoperability and Reusability) of published open data on these substances and to qualitatively evaluate to what extend the selected databases already fulfil the criteria set out in the commission draft regulation on a common data chemicals platform[1].

The FAIR principles have emerged as a guiding framework for ensuring the accessibility and usability of scientific data across disciplines, including chemistry. As a result, the development of tools to assess the FAIRness of data has become increasingly important. These tools can be categorized into two main approaches: manual and automated. Manual tools, such as the Data Archiving and Networked Services (DANS) SATIFYD and the ARDC FAIR Data Self Assessment Tool, are accessible in the form of online questionnaires. These questionnaires provide a structured approach to evaluating FAIRness by guiding users through a series of questions related to the FAIR principles. They are particularly useful for initiating discussions on FAIR implementation within research teams and for identifying areas that require further attention. Automated tools for FAIRness assessment, such as F-UJI and FAIR Checker, are gaining prominence and are continuously under development. Unlike manual tools, automated tools perform a series of tests automatically starting from a dereferenceable URL to the data resource to be evaluated. This approach offers several advantages, including scalability, efficiency, and reduced human effort. Automated tools can assess large datasets effectively and provide a comprehensive overview of their FAIRness. They are particularly valuable for large-scale data initiatives and for ensuring the consistency of FAIR practices across research groups. In this work we will use two automatic FAIRness assessment tools: FAIR Checker and F-UJI.

FAIR-Checker is a free online tool that evaluates whether a dataset meets the FAIR principles for data sharing. To check a dataset, only a valid persistent identifier (PID) or URL is needed to be provided. The tool scans the dataset's landing page and provides a comprehensive assessment of its FAIRness. The results are presented in a visually appealing radar chart, indicating normalized evaluation scores (ranging from 0 for "do not satisfy the principle" to 100 for "completely satisfy the principle") for each FAIR principle. Additionally as an outcome, a table provides detailed scores, test results, log messages, and recommendations for each individual test.
It's important to note that FAIR-Checker's tests don't differentiate between data and metadata evaluations. Concequently, if only the dataset is available, it searches for data description inside of the dataset (e.g., provenance, training material, license, etc.).

---

[1] eur-lex.europa.eu/legal-content/EN/TXT/PDF/?uri=CELEX:52023SC0855

F-UJI is a user-friendly tool that automatically assesses the FAIRness of research datasets. All you need to check a dataset is to provide its URL or a DOI. If there is an online metadata, then you can also provide the URL and specify its type, such as OAI-PMH, OGC CSW, or SPARQL. The assessment process is straightforward and accompanied by progress messages that indicate which maturity indicator is being evaluated. The outcome is a comprehensive report summarizing the assessment results. The report includes a multi-level pie chart visualizing the overall FAIRness level of the dataset. While the pie chart provides a general overview, it's not interactive, meaning you can't click on it to further explore the specific details. The detailed report delves into each test performed, indicating its corresponding FAIR level (initial, moderate, or advanced) using colored checkmarks (light, medium and dark, respectively). Additionally, debug messages are available for every test, allowing users to independently verify and evaluate the test outputs.

The manual analysis of accessibility of datasets is based on two criteria: 1. their presence in FAIRsharing and 2. the availability of links to their datasets via PubChem. If one of these criteria is satisfied, we will apply the following rule for the automatic assessment: The provided URL will be the FAIRsharing link (preferred), or the PubChem link (second option). However, if none of them is available, the URL of the data publisher (where the dataset can be accessed/downloaded) is used.

FAIRsharing is a community-driven resource with users and collaborators across all disciplines that work together to enable the FAIR Principles and promote the use of standards, databases and policies. The goals of FAIRsharing are:
- Classify and align research data policies across publishers and funders
- Moderate cross-publishers discussion on repositories
- Define and register FAIR maturity indicators and metrics
- Build guidance and training material on standards, databases and policies

PubChem is the largest open-access chemistry database, covering hundreds of millions of compounds, substances, and bioactivities. Hosted by the NIH (National Institute of Health), it offers a wealth of chemical information, including physicochemical properties, biological activities, safety and toxicity information, patents, literature citations, and more, connecting almost 1000 data sources. PubChem has links to and descriptions of the sources used to compose their database, expressed in a FAIR format.

## Selected databases

Disclaim:
All the solutions presented here were selected after verifying that public data were in practice findable by anyone on any web browser, since we excluded 'private' solutions during the selection process.

We aimed at defining the inclusion and exclusion criteria as clearly as possible, to make the selection process as transparent as possible. However, because the review was conducted in an academic context, it is possible that some relevant results were missed.
Therefore, we do not claim that the list of solutions provided here is complete, but we believe it is representative of the current data sharing landscape for substance databases involving chemicals data relevant for risk assessment and management.

Selection criteria:

For the purpose of this study, we selected ten freely accessible and widely used chemicals data sources relevant for consumer products and human health risk assessment. The intention was not to be exhaustive within the data source selection, but rather to identify recurrent problems from a selection of widely used sources and contribute to solve them. All chosen data sources respect this set of criteria:

1. The dataset should be widely adopted by researchers, policy makers or public authorities acting on chemical regulation, risk assessment and mitigation within North America and Europe.
2. The dataset is publicly accessible to anyone through any internet browser.
3. The dataset is hosted either in North America or in Europe (for the purpose of scoping of this study. These geographical areas have been active in chemicals data management at the Organisation for Economic Co-operation and Development, OECD, level).
4. The dataset is relevant for consumer exposure and public health (datasets focused purely on environmental health were not considered).

The selected data sources and their Uniform Resource Locators (URLs) are presented in **Table 1** and further described and discussed within the next chapter. Note that at the time of writing this document, the new ECHA CHEM[2] portal was still under development and thus this portal was not taken into consideration for the purpose of this work. For this paper, and for each data source, we applied the two approaches (manual and automatic FAIRness evaluation).

| **Database/platform** | **Accessible URL** |
|---|---|
| European Chemicals Agency (ECHA) REACH registered substance factsheets | https://echa.europa.eu/information-on-chemicals/registered-substances |
| European Chemicals Agency (ECHA) Classification and Labelling (C&L) Inventory | https://echa.europa.eu/information-on-chemicals/cl-inventory-database |
| European Chemicals Agency (ECHA) database for information on **S**ubstances of **C**oncern **I**n articles as such or in complex objects (**P**roducts) (SCIP) | https://echa.europa.eu/scip-database |
| European Commission's Cosmetic ingredient database (Cosing) | https://ec.europa.eu/growth/tools-databases/cosing/ |
| Joint Research Centre's Information platform for chemical monitoring (IPChem) | https://ipchem.jrc.ec.europa.eu/ |
| ChemSpider (Royal Society of Chemistry) | http://www.chemspider.com/ |
| ChemView | https://www.epa.gov/assessing-and-managing-chemicals-under-tsca/introduction-chemview |
| Comparative Toxicogenomics Database (CTD) | https://ctdbase.org/ |
| U.S. Environmental Protection Agency's (EPA) CompTox Chemicals Dashboard | https://comptox.epa.gov/dashboard/ |
| Toxin and Toxin Target Database (Toxic Exposome Database, T3DB) | http://www.t3db.ca/ |

*Table 1 The selected ten data sources as well as the URLs to access the respective source.*

---

[2] https://chem.echa.europa.eu/

# FAIR Analysis

1. ***ECHA REACH***

Description:
   *Website*: https://echa.europa.eu/en/information-on-chemicals/registered-substances
   *FAIRsharing*: Not Available (N.A.)
   *PubChem*: N.A.
   *License*: Free of copyrights
   *API*: N.A.
   *Short explanation*: ECHA's REACH (Regulation (EC) 1907/2006) registered substance list provides a listing of substances that have been registered in the EU by the industry. This table includes all public data submitted to ECHA in REACH registration dossiers by substance manufacturers, importers, or their representatives, as specified in the REACH . ECHA makes this information available as REACH registered substance factsheets.

Analysis:
Using the ECHA URL, the FAIRness score is low since there is no metadata in a format that can be evaluated by FAIR checking tools. The figures below show the results of two automatic FAIR analysis tools: FAIR Checker and F-UJI. The findability score is high because the data can be found in the publisher website and it is downloadable.

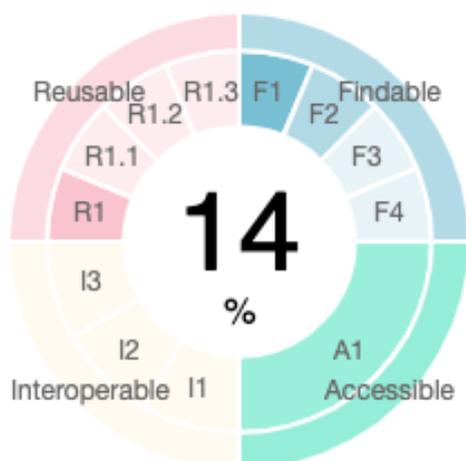 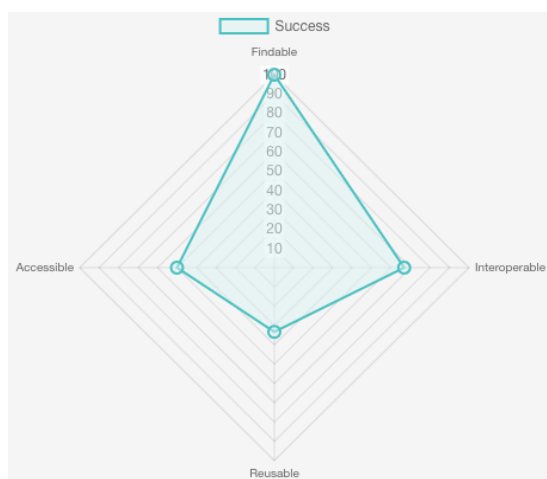

ECHA URL – FAIR analysis (F-UJI)      ECHA URL – FAIR analysis (FAIR Checker)

The outcomes of the manual analysis, together with analysis of the log-summary provided by the two tools resulted in the following findings:
*Findable:*
- REACH registered substance has a permanent and stable web address.

*Accessible:*
- REACH registered substance is accessible from a stable, permanent web address.
- REACH registered substance is accessible by search/browse querying mechanisms.
- REACH registered substance can be downloaded in a machine-readable (CSV, XML, Excel) file format.
- REACH registered substance can be accessed via downloadable files.

*Interoperable:*
- REACH registered substance incorporates terms and IDs (CAS, EC Number).

*Reusable:*

- REACH registered substance data is reusable via the downloadable files.
- REACH registered substance data is reusable in machine-readable formats (CSV, XML, Excel).

Notice:
The database is accessible via the web browser and can be downloaded. The metadata could not be found in the website, thus the automatic FAIRness analysis was limited. The ID are not unique, some substances have CAS number, but not EC Number, and vice-versa. We could find different substances with the same CAS number. Some mix of substances has not CAS number neither EC number.

Suggestions:
- REACH registered substance metadata needs to be provided in standard format (JSON-LD, Dublin Core, RDFa).
- Build the resource ID with a namespace that can be found in Identifiers.org to ensure a persistent identifier accessible by the community.
- REACH registered substance dataset should be registered and indexed in dedicated websites (e.g., FAIRsharing).
- The access via APIs should be implemented.
- Links to websites that give more details about the substance/components should be added (e.g., PubChem, Wikipedia, etc.) and maintained.

2. **C&L Inventory:**

Description:
   *Website*: https://echa.europa.eu/en/information-on-chemicals/cl-inventory-database
   *FAIRsharing*: N.A.
   *PubChem*: N.A.
   *License*: Free of copyrights
   *API*: N.A.
   *Short explanation*: This database contains classification and labelling (C&L) information on notified and registered substances received from manufacturers and importers. It also includes the list of harmonized classifications.

Analysis:
Using the ECHA URL, the FAIRness results are low since there is no metadata in a format that can be evaluated by FAIR checking tools. The figures below show the results of two automatic FAIR analysis tools: FAIR Checker and F-UJI. The findability is high because the data can be found in the publisher website and it is downloadable.

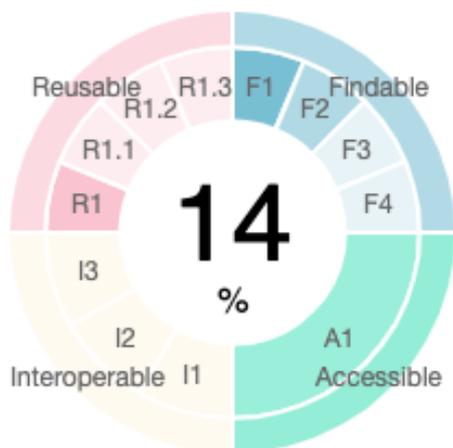 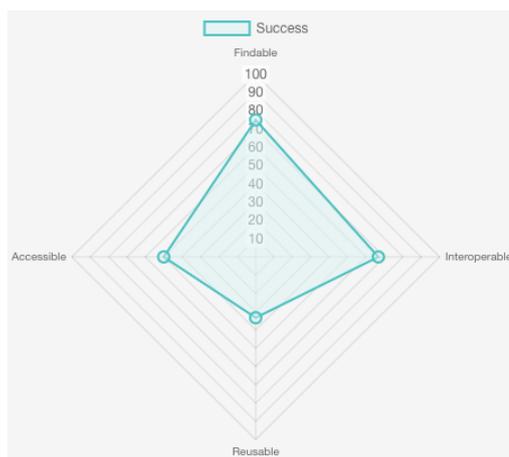

ECHA URL – FAIR analysis (F-UJI)  ECHA URL – FAIR analysis (FAIR Checker)

The outcomes of the manual analysis, together with analysis of the log-summary provided by the two tools resulted in the following findings:

*Findable:*
- C&L Inventory has a permanent and stable web address.

*Accessible:*
- C&L Inventory is accessible from a stable, permanent web address.
- C&L Inventory provides "Training Materials".
- C&L Inventory is accessible by search/browse querying mechanisms.
- C&L Inventory can be downloaded in a machine-readable (CSV, Excel) format.
- C&L Inventory is accessible by downloadable files.

*Interoperable:*
- C&L Inventory incorporates terms and IDs (CAS, EC Number).

*Reusable:*
- C&L Inventory data is reusable via downloadable files.
- C&L Inventory data is reusable in machine-readable formats (CSV, Excel).

Notice:
The database is accessible via the web browser and can be downloaded. The metadata could not be found, and the automatic FAIRness analysis was limited.

Suggestions:
- C&L Inventory metadata should be provided in standard format (JSON-LD, Dublin Core, RDFa).
- Build the resource ID with a namespace that can be found in *Identifiers.org* to ensure a persistent identifier accessible by the community.
- C&L Inventory nshould be registered and indexed in dedicated websites (e.g., FAIRsharing).
- The access via APIs should be implemented.
- Links to websites that give more details about the substance/components should be added (e.g., PubChem, Wikipedia, etc.) and maintained.
- Links to regulations that apply to the substances, if applicable.

### 3. SCIP Database:

Description:
  *Website*: https://echa.europa.eu/scip-database

*FAIRsharing*: N.A.
*PubChem*: N.A.
*License*: Free of copyrights (for the publicly available data only)
*API*: N.A.
*Short explanation*: SCIP is the database for information on Substances of Concern In articles as such or in complex objects (Products) established under the Waste Framework Directive (WFD). Companies supplying articles containing substances of very high concern (SVHCs) on the Candidate List in a concentration above 0.1% weight by weight (w/w) on the EU market are required to submit information on these articles to ECHA, as from 5 January 2021. The SCIP database ensures that the information on articles containing Candidate List substances is available throughout the whole lifecycle of products and materials, including at the waste stage. The information in the database is then made available to waste operators and consumers.

Analysis:
Using the ECHA URL, the FAIRness score is low since there is no metadata in a format that can be evaluated by FAIR checking tools. The figures below show the results of two automatic FAIR analysis tools: FAIR Checker and F-UJI. The findability is high because the data can be found in the publisher website and it is downloadable.

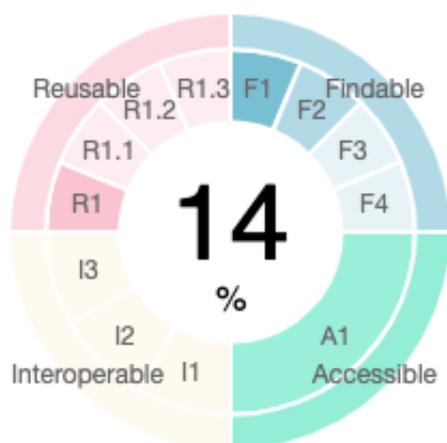
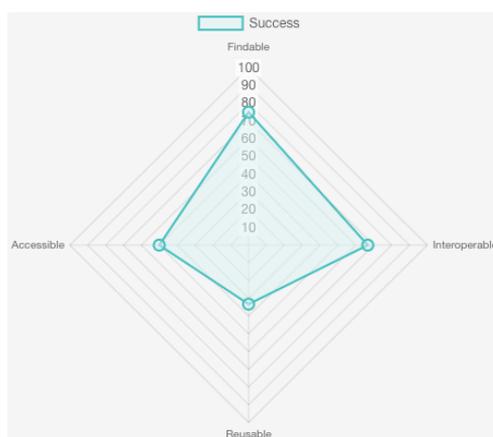

ECHA URL – FAIR analysis (F-UJI)     ECHA URL – FAIR analysis (FAIR Checker)

The outcomes of the manual analysis, together with analysis of the log-summary provided by the two tools resulted in the following findings:

*Findable:*
- SCIP has a permanent and stable web address.

*Accessible:*
- SCIP substance is accessible from a stable, permanent web address.
- SCIP substance is accessible by search/browse querying mechanisms.
- SCIP substance (outcome of queries) can be downloaded in a machine-readable (CSV, XML, Excel) file format.
- SCIP substance can be accessed via downloadable files.

*Interoperable:*
- SCIP substance incorporates terms and IDs (CAS, EC Number).

*Reusable:*
- SCIP substance data is reusable via the downloadable files.
- SCIP substance data is reusable in machine-readable formats (CSV, XML, Excel).

Notice:
The database is accessible via the web browser and the results of the queries can be downloaded. The metadata could not be found on the website, thus the automatic FAIRness analysis was limited. The ID of the substances is not unique, some substances have CAS number, but not EC Number, and vice-versa. We could find different substances with the same CAS number. Some mix of substances have no CAS number neither an EC number.

Suggestions:
- SCIP substance metadata should be provided in standard format (JSON-LD, Dublin Core, RDFa).
- Build the resource ID with a namespace that can be found in *Identifiers.org* to ensure a persistent identifier accessible by the community.
- SCIP dataset should be registered and indexed in dedicated websites (e.g., FAIRsharing).
- The access via APIs should be implemented.
- Links to websites that give more details about the substance/components should be added (e.g., PubChem, Wikipedia, etc.) and maintained.

### 4. COSING:

Description:
*Website*: https://ec.europa.eu/growth/tools-databases/cosing/
*FAIRsharing*: N.A.
*PubChem*: N.A.
*License*: Free of copyrights
*API*: N.A.
*Short explanation*: CosIng is an information-only database that provides a distinction between ingredients and substances. Via the website, you can search for the name of a substance as it is referred to in the Cosmetics Regulation (Regulation (EC) 1223/2009) or for the name of an ingredient and/or fragrance listed in the inventory for labelling purposes. CosIng allows also users to search for relevant CAS and EC numbers.

Analysis:
Using the europa.eu URL, the FAIRness score is low since there is no data and metadata in a format that can be evaluated by FAIR checking tools. The figures below show the results of two automatic FAIR analysis tools: FAIR Checker and F-UJI. The findability looks high because the URL of the publisher is a permanent one.

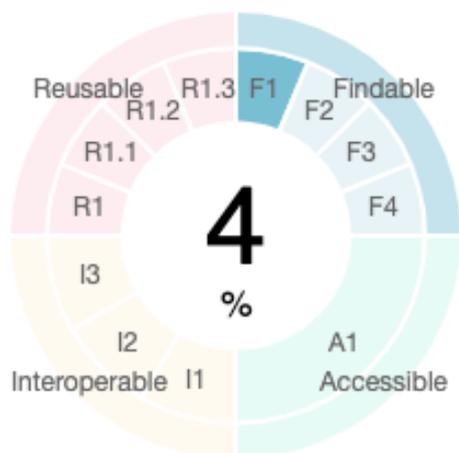 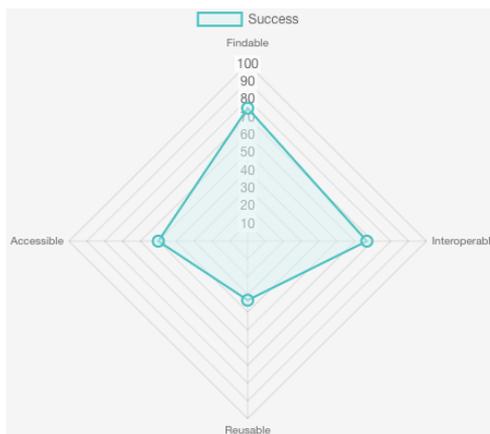

ECHA URL – FAIR analysis (F-UJI)　　　ECHA URL – FAIR analysis (FAIR Checker)

The browser available in the website of Cosing allows querying for substances, however it is not possible to download the list of substances, neither the outcome of the query, in a computer-interpretable format. Following the links on the website, it is possible to access PDF files describing each substance. This explains the low FAIRness score for Cosing. The outcomes of the manual analysis, together with analysis of the log-summary provided by the two tools resulted in the following findings:

*Findable:*
- CosIng has a permanent and stable web address for browsing.

*Accessible:*
- CosIng's substances are accessible from a stable, permanent web address.
- CosIng's substances are accessible by search/browse querying mechanisms.

*Interoperable:*
- CosIng incorporates terms and IDs (CAS, EINECS, ELINCS, INCI).

*Reusable:*
- CosIng data is not available for reuse.

Suggestions:
- CosIng datashould be downloadable in a standard format (e.g., CSV, XML, JSON).
- CosIng metadata should be provided in standard format (JSON-LD, Dublin Core, RDFa).
- CosIng's metadata should be registered and indexed in dedicated websites (e.g., FAIRsharing).
- The provenance of the datashould be included in the dataset. Explicitly indicate the EU regulation that prompted the the information gathering (or any other source used to create the dataset)
- The access via APIs should be implemented.
- Links to websites that give more details about the substance/ingredient should be added (e.g., PubChem, Wikipedia, etc.) and maintained.
- Links to regulations that applies to the substances, if applicable.

### 5. IPCHEM:
Description:
　　*Website*: https://ipchem.jrc.ec.europa.eu/

*FAIRsharing*: N.A.
*PubChem*: https://pubchem.ncbi.nlm.nih.gov/source/11946
*License*: N.A.
*API*: N.A.
*Short explanation:* The Information Platform for Chemical Monitoring is the European Commission's reference access point for searching, accessing and retrieving chemical occurrence data collected and managed in Europe. The platform has been developed to fill the knowledge gap on chemical exposure and its burden on health and the environment. IPCHEM is structured into four modules, according to the chemical monitoring data categorization: Environmental monitoring, Human Bio-Monitoring, Food and Feed, Products and Indoor Air. The role of IPCHEM is to become a registry for chemical data and to provide one central entering point for querying several datasets. The outcome will be the list of links to datasets where the searching information was found.

Since IPCHEM is not a dataset (but an index), the automatic FAIR analysis failed. The manual analysis produced the following results:
*Findable:*
- IPCHEM has a unique identifier (the URL).

*Accessible:*
- IPCHEM is accessible from a stable, permanent web address.
- IPCHEM provides an "Training Materials".
- IPCHEM provides search/browse querying mechanisms.

*Interoperable:*
N.A.
*Reusable:*
N.A

Notice:
A small set of datasets are registered in IPCHEM. The website was developed to be used by humans and it is not yet suitable for automatic actions to access and reuse the data.

Suggestions:
- Extend the registry to aggregate more data.
- Provide an API to automatize the searching and the accessing processes.
- Publish the metadata of datasets required to automatically access and interpret the data, promoting the interoperability between systems and the reusability of data.
- Use standard vocabularies for metadata description (including license and provenance).
- Develop a visual searching tool that facilitates discovering data/values and exploring the relationships between them.

6. ***ChemSpider:***

Description:
   *Website*: https://www.chemspider.com/
   *FAIRsharing*: https://doi.org/10.25504/FAIRsharing.96f3gm
   *PubChem*: https://pubchem.ncbi.nlm.nih.gov/source/ChemSpider
   *License*: Free for research use
   *API*: https://developer.rsc.org/apis

*Short explanation:* ChemSpider is a free chemical structure database providing fast text and structure search access to over 100 million structures from hundreds of data sources. In this database, you can also find literature references, physical properties, spectra, chemical suppliers, etc. The Royal Society of Chemistry also provides an API built on top of their ChemSpider to access the data.

Using the ChemSpider URL, the FAIRness is low (~22%) since the metadata is incomplete, according to the criteria of the FAIR checking tools. However, the FAIRsharing community added some metadata to ChemSpider and stored it in the registers of FAIRsharing. The figures below show the results of two automatic FAIR analysis tools: FAIR Checker and F-UJI.

Analysis:
For this dataset, we want to show the difference between using the publisher URL and the FAIRSharing URL. With the publisher URL, the FAIRness score is lower since the metadata is not complete or not in a format that can be evaluated by FAIR checking tools. The figures below shows the results (for each URL) of two automatic FAIR analysis tools: FAIR Checker and F-UJI.

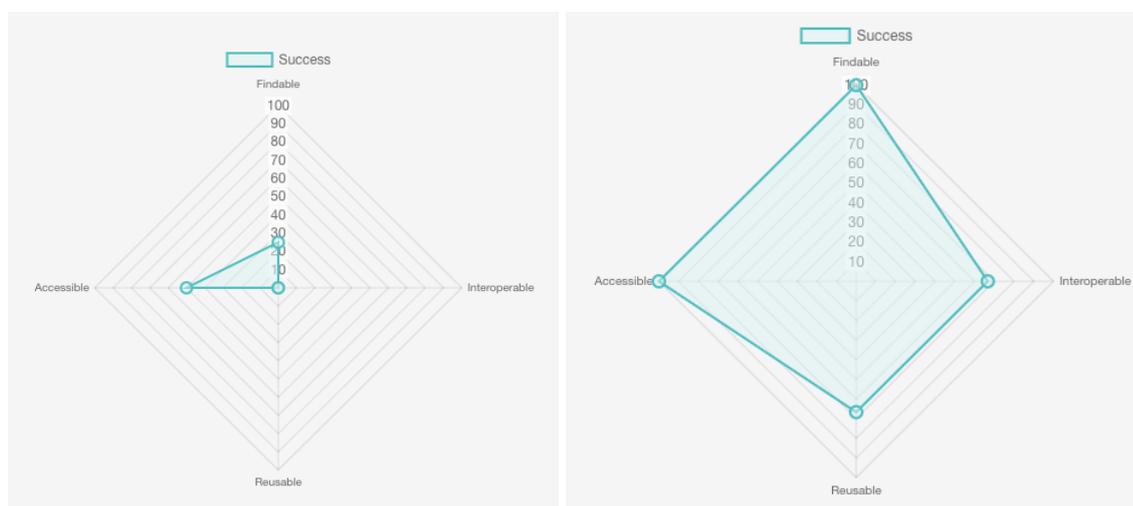

ChemSpider URL – FAIR analysis (FAIR Checker)     FAIRsharing URL – FAIR analysis

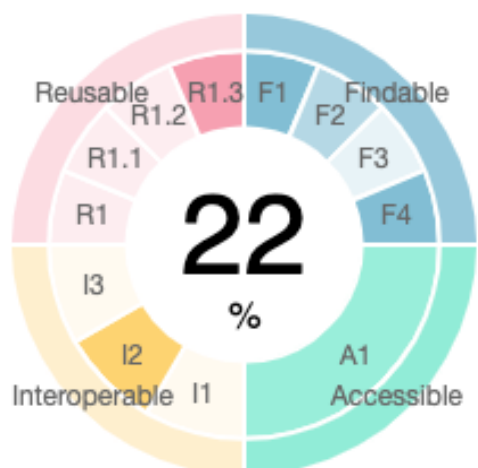 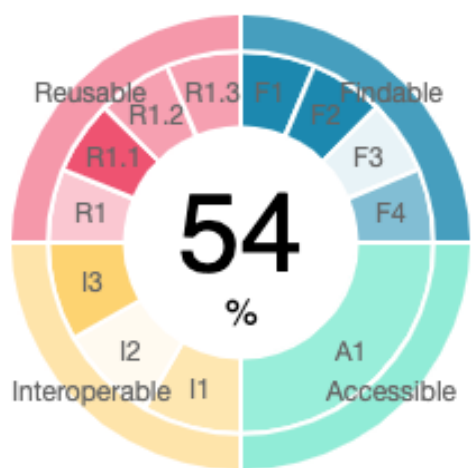

ChemSpider URL – FAIR analysis (F-UJI)             FAIRsharing URL – FAIR analysis

The outcomes of the manual analysis, together with analysis of the log-summary provided by the two tools resulted in the following findings:

*Findable:*
- ChemSpider has a permanent and stable web address.
- ChemSpider entries are uniquely identified by stable URLs.
- ChemSpider is indexed in the FAIRsharing repository. It has a persistent ID in FAIRsharing.
- Metadata is not published on their website in a machine understandable format. But the file registered in FAIRsharing can be used as metadata.

*Accessible:*
- ChemSpider is accessible from a stable, permanent web address.
- ChemSpider provides an "Training Materials".
- ChemSpider can be queried using a web-based form.
- ChemSpider is accessible by search/browse querying mechanisms.
- ChemSpider is accessible via an API tools.

*Interoperable:*
- ChemSpider: incorporates terms, synonyms, and IDs.
- ChemSpider has links to Wikipedia, PubChem, and many other websites.

*Reusable:*
- ChemSpider data provenance is provided, but not in PROV-O.
- ChemSpider data-usage license is available on the website, but not in the metadata.

Notice:
- The information in FAIRsharing is not an official information provided and maintained by US EPA. It can contain outdated information.
- Not downloadable integrally, not available for reuse (except for personal use)

Suggestions:
- The format of the metadata to be published on the website should follow the most used ones (JSON-LD, Dublin Core, RDFa).
- Although a minimal set of information about the data is provided in the FAIRsharing metadata, more details are necessary. For instance, the structure, fields descriptions, etc.
- The metadata should explicitly indicate links to other data sources (this information is in the dataset, but not in the metadata).
- The information about the provenance is provided, but not in a standard format (PROV-O).
- The access conditions should be included in the metadata.

### 7. ChemView

Description:
   *Website*: https://chemview.epa.gov/chemview
   *FAIRsharing*: N.A
   *PubChem*: N.A.
   *License*: Free of copyright
   *API*: https://chemview.readthedocs.io/en/stable/quick.html
   *Short explanation:* ChemView is part of US EPA's commitment to strengthen its chemicals management programs by improving access to and the usefulness of chemical information.

ChemView provides key information in a layered summary format and provides links to underlying studies or other source documents. At this time, users can find information organized in templates for the following: (1) Data Submitted to EPA; (2) EPA Assessments; (3) EPA Actions; (4) Manufacturing, Processing, Use, and Release Data.

Analysis:
According to the rules to choose the URL that will be used in this analysis, we selected the FAIRSharing one. The figures below show the results of two automatic FAIR analysis tools: FAIR Checker and F-UJI. ChemView is a data aggregator and not a dataset per se. The analysis tools could not find enough information about the source data that allows better score. A manual analysis complement this work, as reported below:

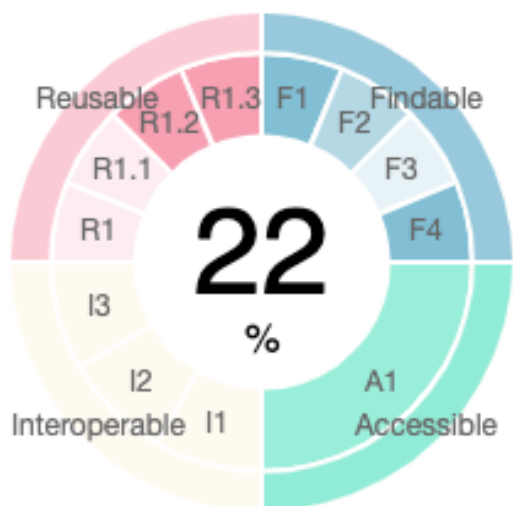
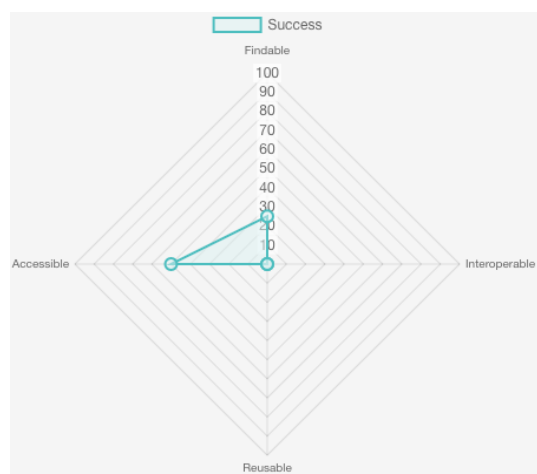

ChemView URL – FAIR analysis (F-UJI)     ChemView URL – FAIR analysis (FAIR Checker)

*Findable:*
- ChemView has a stable web address.
- ChemView entries are uniquely identified by stable URLs.
- Metadata is not published in a machine understandable format.
- ChemView metadata is not yet registered in FAIRSharing or PubChem.

*Accessible:*
- ChemView is accessible from a stable web address.
- ChemView provides a "Documentation" explaining the content of the database.
- ChemView downloaded data is machine-readable (Excel).
- ChemView is accessible by search/browse querying mechanisms.
- ChemView content is accessible by downloadable files.

*Interoperable:*
- ChemView incorporates terms and IDs (CAS, PNM).

*Reusable:*
- ChemView data is reusable in downloadable files.
- ChemView data is reusable in machine-readable formats (Excel).

Suggestions:
- The URL of the datasets should be in the metadata.
- The vocabularies used to produce the metadata should follow the standards (include also the namespaces).

- Registry the ID of the database in at least one registry from the list of persistent registries
- The provenance of the data should also be indicated in the metadata.
- The format of the metadata to be published in the website needs to follow the most used ones (JSON-LD, Dublin Core, RDFa).
- The license to use the data should be explicitly indicated in the metadata.
- Give detailed description of the data.

## 8. CTD

Description:
    *Website*: https://ctdbase.org/
    *FAIRsharing*: https://doi.org/10.25504/FAIRsharing.h3tjtr
    *PubChem*:
    https://pubchem.ncbi.nlm.nih.gov/source/Comparative%20Toxicogenomics%20Database
    *License*: Reproduction or use for commercial purpose is prohibited
    *API*: N.A.
    *Short explanation:* CTD is a robust, publicly available database that aims to advance understanding about how environmental exposures affect human health. It provides manually curated information about chemical–gene/protein interactions, chemical–disease and gene–disease relationships. These data are integrated with functional and pathway data to aid in development of hypotheses about the mechanisms underlying environmentally influenced diseases.

Analysis:
Using the FAIRSharing URL gives a good score of FAIRness to CTD. The figures below show the outcome of the analysis done by both tools FAIR Checker and F-UJI.

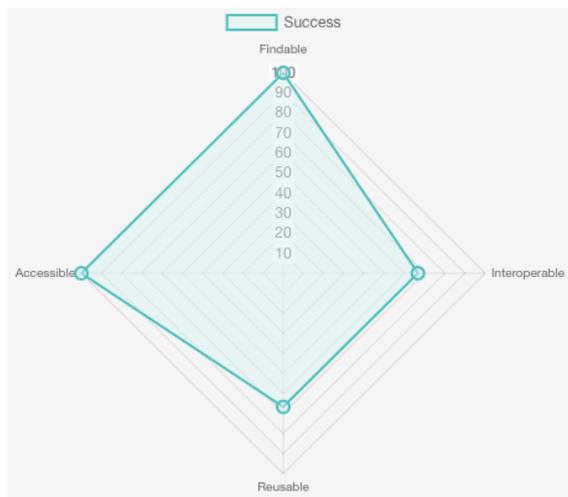
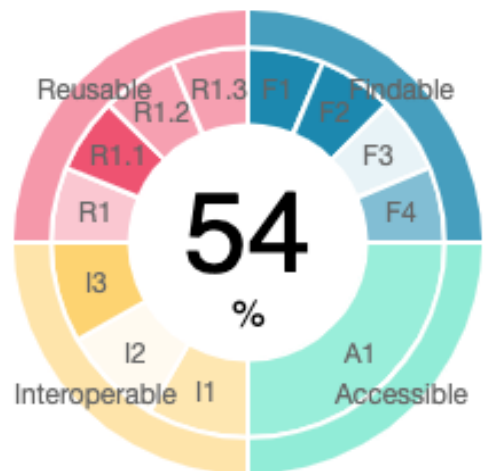

FAIRsharing URL – FAIR analysis (FAIR Checker)    CTD URL – FAIR analysis (F-UJI)

CTD was the only publisher that self-analysed the FAIRness of their dataset. We decided to show in this document this analysis and comment (grey text) on it, if needed. We add our suggestions of improvements below their analysis.

*Findable:*

- [CTD is findable from a stable, permanent web address](). (it is also indexed in PubChem and FAIRSharing)
- CTD entries are uniquely identified by stable URLs

*Accessible:*
- CTD is freely accessible, and does not require account creation or logon.
- [CTD is accessible from a stable, permanent web address]().
- [CTD data can be linked to following these instructions]().
- [CTD provides an "Introduction" video tutorial]() and [online "Help" guide]().
- [CTD can be contacted using a web-based form](), logged via Request Tracker system.
- CTD downloaded data is machine-readable (JSON, XML, XSD, CSV, TSV, Excel).
- [CTD is accessible by search/browse querying mechanisms]().
- [CTD is accessible by downloadable files]().

*Interoperable:*
- The reuse of external identifiers is a good practice and facilitates the connection between data sources
- [CTD Chemicals](): defined by unique [MESH:ID]()
- [CTD Genes](): defined by unique [NCBI Gene:ID]()
- [CTD Phenotypes](): defined by unique [Gene Ontology:ID]()
- [CTD Diseases](): defined by unique [MESH:ID]() or [OMIM:ID]()
- [CTD Anatomy](): defined by unique [MESH:ID]()
- [CTD Organisms](): defined by unique [NCBI Taxonomy:ID]()
- [CTD References](): defined by unique [PubMed:PMID]()
- [Gene Ontology annotations](): defined by unique [Gene Ontology:ID]()
- [Pathway annotations](): defined by unique [KEGG:ID]() or [Reactome:ID]()

*Reusable:*
- CTD data provenance is provided with relevant attributes. (not in the metadata)
- [CTD data-usage license is available](). (not in the metadata)
- CTD curated data is traceable to original source material (via PMID).
- [CTD curation paradigms are fully described and transparent in CTD publications]().
- [CTD curation terms are defined]().
- [CTD changes and updates are logged and displayed]().
- [CTD data is reusable in downloadable files]().
- CTD data is reusable in machine-readable formats (JSON, XML, XSD, CSV, TSV, Excel).

Suggestion:
CTD worked to improve the FAIRness of their data and listed all the criteria/actions to reach this goal. As a suggestion, CTD can publish in their website the metadata in a standard format that describes the dataset. It is also necessary to publish the license information in the metadata. The metadata has partially been created and published in FAIRsharing registry and can be reused by the CTD team.

9. *Comptox*

Description:
   *Website*: https://www.epa.gov/comptox-tools
   *FAIRsharing*: https://doi.org/10.25504/FAIRsharing.tfj7gt
   *PubChem*: N.A.
   *License*: Free of copyrights

*API*: https://www.epa.gov/comptox-tools/computational-toxicology-and-exposure-data-apis

*Short explanation:* The CompTox Chemicals Dashboard is a part of a suite of databases and web applications developed by EPA. These databases and apps support EPA's research efforts to develop and apply new approach methods (NAMs). EPA researchers integrate advances in biology, bioinformatics, biotechnology, chemistry, and computer science to identify important biological processes that may be disrupted by chemicals. The combined information helps prioritize chemicals based on potential health risks. Using computational toxicology research methods, thousands of chemicals can be evaluated for potential risk at small cost in a very short amount of time. The APIs provided by EPA enable users to extract specific data from various databases and integrate them into their applications. These data are also available for download on their Data Download page : https://www.epa.gov/comptox-tools/downloadable-computational-toxicology-data

Analysis:
Using the FAIRSharing URL gives a good score of FAIRness to Comptox. The figures below show analysis done by both tools FAIR Checker and F-UJI.

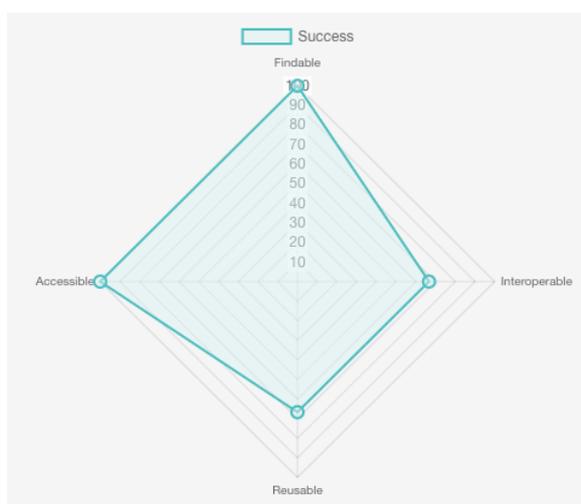 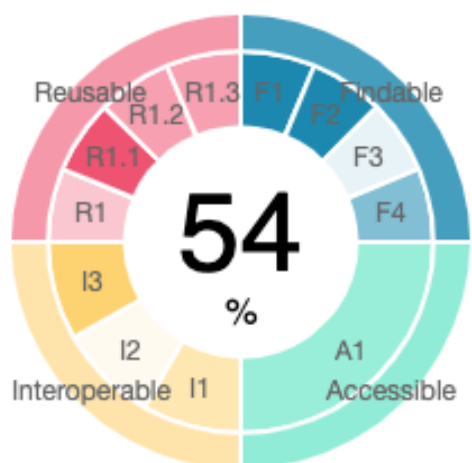

EPA URL – FAIR analysis (FAIR Checker)       EPA URL – FAIR analysis (F-UJI)

*Findable:*
- Comptox has a stable web address, managed by a governmental institution (EPA)
- Comptox entries are uniquely identified by stable URLs.
- Comptox is indexed in the FAIRsharing repository, but not yet in the PubChem. It has a persistent ID in FAIRsharing.
- Metadata is not published in a machine understandable format on the website of Comptox. However, the file registered in FAIRsharing can be used as metadata.

*Accessible:*
- Comptox is accessible from a stable, permanent web address.
- Comptox provides "Training Materials".
- Comptox downloadable data is machine-readable (CSV, SDF, Excel).
- Comptox data is accessible by search/browse querying mechanisms.
- Comptox data is accessible via an API tools.
- Comptox data is accessible in the downloadable files.

*Interoperable:*

- Comptox Chemicals: incorporates terms, synonyms, and IDs (CAS, DTXSID).
- Comptox has links to Wikipedia, Chebi, PubChem, and many other websites.

*Reusable:*
- Comptox data provenance is provided in the dataset, but PROV-O is not adopted.
- Comptox data-usage license is available on the website, but not in the metadata.
- Comptox indicates the curation level in the dataset.
- Comptox use publisher APIs to find related publications.
- Comptox changes and updates are logged and displayed in the website.
- Comptox data is reusable in downloadable files.
- Comptox data is reusable in machine-readable formats (CSV, SDF, Excel).
- Several tools using Comptox data is available.

Notice:
- The information in FAIRsharing is not an official information provided and maintained by EPA. It can contain outdated information.

Suggestions:
- The format of the metadata can follow the most used ones (JSON-LD, Dublin Core, RDFa).
- The namespaces of semantic resources need to be included in the metadata or it should be used the one from identifiers.org.
- Although a minimal set of information about the data is provided, the metadata could detail more the data.
- The metadata should explicitly indicate links to other data sources (this information is in the dataset, but not in the metadata).
- The information about the provenance is provided, but not in a standard format (PROV-O).
- The access conditions need to be included in the metadata.
- The link to the database should be added to the metadata.

### 10. T3DB (Toxic Exposome Database) :

Description:
 Website: http://www.t3db.ca/
 FAIRsharing: https://doi.org/10.25504/FAIRsharing.68a369
 PubChem: https://pubchem.ncbi.nlm.nih.gov/source/25803
 License: Free of copyright
 API: N.A.
 Short explanation: The *Toxin and Toxin Target Database (T3DB)*, also referred as, the *Toxic Exposome Database*, is a unique bioinformatics resource that combines detailed toxin data with comprehensive toxin target information. The database currently houses 3,678 toxins described by 41,602 synonyms, including pollutants, pesticides, drugs, and food toxins, which are linked to 2,073 corresponding toxin target records. Altogether there are 42,374 toxin, toxin target associations. The focus of the T3DB is on providing mechanisms of toxicity and target proteins for each toxin.

Analysis:
Using the FAIRSharing URL gives a good score of FAIRness to T3DB. The figures below show analysis done by both tools FAIR Checker and F-UJI.

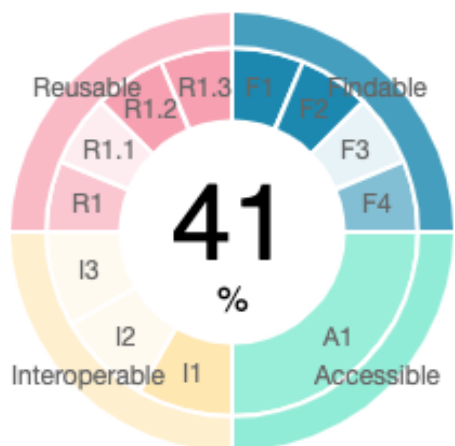 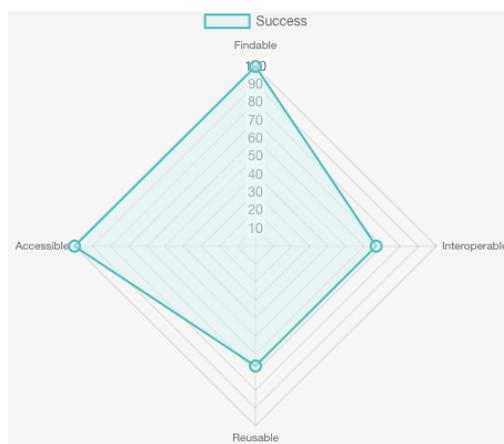

FAIRsharing URL – FAIR analysis (F-UJI)　　FAIRsharing URL – FAIR analysis (FAIR Checker)

*Findable:*
- T3DB has a permanent and stable web address.
- T3DB entries are uniquely identified by stable URLs.
- T3DB is indexed in the FAIRsharing repository and in PubChem. It has a persistent ID.
- Metadata is not published in a machine understandable format in the editor website. But the file registered in FAIRsharing can be used as metadata (but need improvements).

*Accessible:*
- T3DB is accessible from a stable, permanent web address.
- T3DB provides a short "Documentation" about the fields.
- T3DB downloaded data is machine-readable (CSV, XML, JSON, SDF, FASTA).
- T3DB is accessible by search/browse querying mechanisms.
- T3DB is accessible by downloadable files.

*Interoperable:*
- T3DB has terms, synonyms, and IDs (CAS, CAS, IUPAC, InChl, etc.).
- T3DB has links to Wikipedia, Chebi, PubChem, and many other websites. However, these links are not described in the metadata.

*Reusable:*
- T3DB data provenance is provided in the dataset, but not in the metadata.
- T3DB data-usage license is available on the website, but not in the metadata.
- T3DB data is reusable in downloadable files.
- T3DB data is reusable in machine-readable formats (CSV, SDF, Excel).

Suggestions:
- Add license information to the metadata.
- The format of the metadata to be published on the website can follow the most used ones (JSON-LD, Dublin Core, RDFa).
- Although a minimal set of information about the data is provided, the metadata could detail more the data.

## Conclusions

We analysed ten widely adopted datasets managed in Europe and North America. The highest score from automatic analysis was 54/100. The manual analysis shows that several FAIR metrics were satisfied, but not detectable by automatic tools because there is not metadata or the format of the information was not a standard one, so it was not interpretable by the tool. We harmonized the analysis and regrouped the scores into the four categories (0 – the analysis fails because the data/metadata is not accessible, 4 – Good quality of metadata):

- 0 – The metadata could not be found in the publisher website or in the selected indexing repository (i.e., fairsharing.org)
- 1 – A set of relevant information was found and we could compute the FAIRness, however the metadata is not complete, and a subset of important information is missing. This value was selected when the automatic evaluation provide a rate inferior to 25% and the manual is inferior to 50% of compliance.
- 2 - A set of relevant information was found and we could compute the FAIRness, however we suggest improvements to the description, and we propose to complete the metadata with information that facilitates its reuse. This value was selected when the automatic evaluation is rated between 25% and 50%, and the manual evaluation rate is superior to 50%.
- 3 – All important information is available and has good quality, but still need some precisions. This value was selected when the automatic evaluation is rated between 50% and 75%, and the manual evaluation rate is superior to 75%.
- 4 - The quality of the metadata is very good. This value was selected when both rates are superior to 75%.

| Database/platform | Scores | Challenges |
|---|---|---|
| European Chemicals Agency (ECHA) REACH registered substance factsheets | 1 | Add and share metadata<br>Add API<br>Add links to other datasets |
| European Chemicals Agency (ECHA) Classification and Labelling (C&L) Inventory | 1 | Add and share metadata<br>Add API<br>Add links to other datasets |
| European Chemicals Agency (ECHA) database for information on **S**ubstances of **C**oncern **I**n articles as such or in complex objects (**P**roducts) (SCIP) | 1 | Add and share metadata<br>Add API<br>Add links to other datasets |
| European Commission's Cosmetic ingredient database (Cosing) | 1 | Add and share metadata<br>Make data downloadable<br>Add API<br>Add links to other datasets and regulation |
| Joint Research Centre's Information platform for chemical monitoring (IPChem) | 1 | Add and share metadata<br>Add API<br>Add links/metadata of aggregated datasets |
| ChemSpider (Royal Society of Chemistry) | 3 | Use standard format of metadata<br>Details more the metadata<br>- provenance, license, etc. |
| ChemView | 2 | Use standard format of metadata<br>Details more the metadata<br>- provenance, license, etc. |

|  |  |  | - detail the sources |
| --- | --- | --- | --- |
| Comparative Toxicogenomics Database (CTD) |  | 3 | Use standard format of metadata<br>Details more the metadata<br>  - provenance, license, etc. |
| U.S. Environmental Protection Agency's CompTox Chemicals Dashboard |  | 3 | Use standard format of metadata<br>Details more the metadata<br>  - provenance, license, etc. |
| Toxin and Toxin Target Database (Toxic Exposome Database, T3DB) |  | 2 | Use standard format of metadata<br>Details more the metadata<br>  - provenance, license, etc.<br>  - detail the sources |

*Table 2: Summary of scores and challenges*

| **Overall issue** (identified by the Commission) | **Specific issue** (identified by the Commission) | **Suggestions on how this could be addressed** |
| --- | --- | --- |
| Information is difficult to find, or share | Chemicals data is scattered | *The solution would be having "aggregators" that provides one single query point to search information from several datasets* |
|  | Globally unique IDs (GUIDs) are not available or adopted by data publishers | *Standard organizations should provide "free-for-use" IDs that are unique for substances, mixtures, and classes. This IDs should be persistent and available in a standard format (e.g., identifiers.org).* |
|  | Chemicals data is not always interoperable | *Data and metadata with a minimal set of information should be made available by data publishers (at least in official data sources)* |
|  | Chemicals data is not always accessible | *Data should be openly accessible and policies for exceptions must be defined to preserve IP and commercial interests, without impacting on data quality or GUIDs.* |
| Knowledge base is incomplete | Academic data are insufficiently considered | Addressed by the duty to notify within the draft regulation on a common chemicals data platform. |
|  | Lack of availability of certain types of chemicals data | *Researchers should be trained to FAIRly publish their data and rewarded when they reach the* |

|  | | *quality/compliance criteria of "aggregators"* |
|---|---|---|
|  | Not all study results are reported by duty holders | Same as before |
|  | Lack of mechanism to identify emerging chemical risks | Creation of methodologies to identify early warning signs once the issues concerning data FAIRness have been resolved. |

*Table 3 Issues related to chemicals data and summarised by the Commission Staff working document published in December 2023³ as well as solutions proposed based on this FAIRness study.*

Future works focus on deep analyse the European regulations and provide insights on how these regulations can be modified to reinforce the FAIRness of datasets. We will also develop a knowledge base with aggregated data from several sources and demonstrate that recent ICT technologies can be used to better support researchers to collect and exploit chemical data in their studies.

---

[3] Proposal for a REGULATION OF THE EUROPEAN PARLIAMENT AND OF THE COUNCIL Establishing a Common Data Platform on Chemicals, Laying down Rules to Ensure That the Data Contained in It Are Findable, Accessible, Interoperable and Reusable and Establishing a Monitoring and Outlook Framework for Chemicals. https://eur-lex.europa.eu/resource.html?uri=cellar:1cf46b5f-94f6-11ee-b164-01aa75ed71a1.0001.02/DOC_1&format=PDF (accessed 2024-04-11).